\documentstyle[11pt]{article}
\newcommand{\bhbar}{\mbox{$\bar{\Delta}H$ }}
\newcommand{\dh}{\mbox{$\Delta H$ }}
\newcommand{\bwbar}{\mbox{$\bar{\Delta}W$ }}
\begin{document}
\title{Uniaxial Phase Transition in Si: {\em Ab initio} Calculations }
\author{C. Cheng$^{*}$ \\
 Department of Physics, National Cheng Kung University, Tainan, Taiwan, R.O.C.}        
\date{}
\maketitle
\begin{abstract}
Based on a previously proposed thermodynamic analysis$^{\cite{paperI}}$, we study the relative stabilities of five Si phases under unaxial compression using {\em ab initio} methods.
The five phases are diamond, $\beta$Sn, simple-hexagonal (sh), simple-cubic, and hcp structures.
The possible phase-transition patterns were investigated by considering the phase transitions between any two chosen phases of the five phases.
By analyzing the different contributions to the relative phase stability, we identified the most important factors in reducing the phase-transition pressures at uniaxial compression.
We also show that it is possible to have phase transitions occur only when the phases are under uniaxial compression, in spite of no phase transition when under hydrostatic compression.

Taking all five phases into consideration, the phase diagram at uniaxial compression was constructed for pressures up to 20 GPa.  
The stable phases were found to be diamond, $\beta$Sn and sh structures, i.e. the same as those when under hydrostatic condition.
According to the phase diagram, direct phase transition from the diamond to the sh phase is possible if the applied uniaxial pressures, on increasing, satisfy the condition of $P_{x}>P_{z}$. 
Similarly, the sh-to-$\beta$Sn transition \textit{on increasing pressures} is also possible if the applied uniaxial pressures are varied from the condition of $P_{x}>P_{z}$, on which the phase of sh is stable, to that of $P_{x}<P_{z}$, on which the $\beta$Sn is stable.
\end{abstract}
\begin{flushleft}
Classification Number: 05.70.Ce, 61.50.Ks, 64.70.Kb  \\
$*$e-mail : ccheng@phys.ncku.edu.tw  \\
\end{flushleft}

\newpage
\section{Introduction}

Most theoretical studies of compressive phase transitions are restricted to hydrostatic compression.
However, nonhydrostatic compression occurs commonly in realistic situations and can lead to very different phase-transition pressures.
For example, previous experimental studies have shown$^{\cite{si-uniexp1,si-uniexp2}}$ that nonhydrostatic compression effectively reduces the phase-transition pressure of diamond structured Si to the metallic $\beta$Sn phase. 
The generally used enthalpy function $H=E+PV$ in {\em ab initial} approach, which has been very successful in predicting hydrostatic phase-transition pressure$^{\cite{ab1}}$, is no longer the appropriate quantity in determining the relative stability between phases under nonhydrostatic conditions.
In a previous paper$^{\cite{paperI}}$ (called paper I hereafter), we have carried out a thermodynamic quantity analysis, similar to the use of the enthalpy difference in hydrostatic cases, in order to distinguish the relative stability between phases under uniaxial compression.
That was applied to the uniaxial phase transitions between the diamond and $\beta$Sn structures of Si and Ge using {\em ab initio} calculations.
It was found that under uniaxial compression the phase-transition pressures for both Si and Ge were greatly reduced, i.e. from 11.4 to 3.9 GPa for Si and 9.5 to 2.5 GPa for Ge.

In the present work, based on the previous thermodynamic analysis, we study the relative stability of five Si phases under uniaxial compression. 
The five phases are diamond, $\beta$Sn, simple hexagonal (sh), simple cubic (sc) and hexagonal closed packed (hcp) structures.
We investigate the possible phase-transition patterns at uniaxial compression by considering the phase transitions between any two chosen phases of the five structures.
We shall show that it is possible to have phases which do not undergo phase transition when under hydrostatic compression but do undergo phase transition when under uniaxial compression.
By analyzing the different contributions to the relativie phase stability, we identify the most important factors in reducing the phase-transition pressures at uniaxial compression.  
We demonstrate that the zero-pressure structural parameters of two phases can be used to guess the phase-transition pattern of the two phases under uniaxial compression. 
Finally, the phase diagram of the five Si structures under uniaxial compression is presented.

The paper is organized as the followings: in the next section the {\em ab initio} methods used in the studies are specified.  
The third section describes the hydrostatic results.
The fourth section presents the {\em ab initially} obtained phase-transition patterns at uniaxial compression and discusses the important factors leading to different phase-transition patterns.
In the fifth section we present the uniaxial phase diagram of the five Si phases and finally the conclusions in the sixth section.

\section{Calculational Methods}

In our studies, the total energy (with respect to the atomic energies) of systems were calculated with the {\em ab initio} density functional method$^{\cite{DFT}}$, and the generalized gradient approximation (GGA)$^{\cite{PW1}}$ for the exchange and correlation functional was used.  
The calculations were performed using the {\em ab-initio} total-energy and molecular-dynamics program VASP (Vienna {\em Ab-initio} Simulation Program) developed at the Institut f\"{u}r Material Physik of the Universit\"{a}t Wien $^{\cite{VASP1,VASP2,VASP3}}$.
Norm-conserving pseudopotentials$^{\cite{pot}}$ were used for the electron-ion interactions.
All present calculations were carried out in momentum space and the Kohn-Sham wavefunctions$^{\cite{KS}}$ were expanded in a plane-wave basis.
The energy cutoffs for the plane-wave basis was 400eV.
The integration over the Brillouin zone was approximated by k-point sampling using the Monkhorst-Pack method$^{\cite{MP}}$.
For all the five structures considered in this study, we used unit cells consisting of 8 Si atoms.  
The conventional tetragonal cells were used for the diamond and $\beta$Sn structures while for the sh and sc structures, the $2\times2\times2$ cells of the conventional primitive cells were used.
For the hcp structure, we took the $2\times2\times1$ of the conventional primitive cell to make an 8-atom unit cell.
The Monkhorst-Pack parameters listed in Tab. 1 correspond to those used in the respective 1st Brillouin zones of the 8-atom unit cells just described.

The total energies, i.e. $E(a_{lateral},a_{z})$, were evaluated at various lattice constants $a_{lateral}$ and $a_{z}$ of the unit cells for every 0.05 $\AA$.
The ranges of lattice constants included in the calculations were listed in Tab. 1. 
The diagonal components of a uniaxial stress, i.e. $P_{x}=P_{y}$ and $P_{z}$ (we shall for convenience refer to as 'pressures'), were obtained from the partial derivatives of the evaluated total energy surface $E(a_{lateral},a_{z})$, i.e. 
\begin{equation}
 P_{x}=P_{y}=\frac{1}{\frac{2A}{a_{lateral}}a_{z}} \frac{\partial E(a_{lateral},a_{z})}{\partial a_{lateral}}|_{a_{z}} 
 \mbox{ } \mbox{ and } \mbox{ }
P_{z}=\frac{1}{A} \frac{\partial E(a_{lateral},a_{z})}{\partial a_{z}}|_{a_{lateral}}.
\end{equation} 
The $A$ is the area of the lateral unit cell, i.e. $a_{lateral}^2$ for diamond, $\beta$Sn and sc structures and $\frac{\sqrt{3}}{2}a_{lateral}^{2}$ for sh and hcp structures.
Only structures under pressures of less than 20GPa were included in the following discussions.
Calculations with a higher energy cutoff and different k-point set were considered$^{\cite{paperI}}$.
The change in transition pressure due to numerical errors is expected to be less than 0.5 GPa.

\section{Results at hydrostatic condition}

The calculated energy-versus-volume relations for the five structures under hydrostatic condition, i.e. $P_{x}=P_{z}$, were plotted in Fig. 1.
The lattice constants of the previously defined 8-atom unit cells for the five phases at zero pressure were listed in Tab. 2. 
They are in good agreement with the experimental and previous theoretical results$^{\cite{si-exp2,si-gga}}$.
Within the pressure ranges we considered in this study, i.e. $P_{x}<20GPa$ and $P_{z}<20GPa$, the stable phases were found to be diamond, $\beta$Sn and sh structures.
The hydrostatic transition pressures (call $P_{hydro}$ hereafter) are 11.4 GPa and 13.7 GPa for the diamond-to-$\beta$Sn and $\beta$Sn-to-sh transitions respectively.
These are consistent with previous studies$^{\cite{si-exp2,si-gga}}$.
In the present studies, we did not include a phase of space group {\em Imma} which have been identified, both experimentally$^{\cite{imma-exp}}$ and theoretically$^{\cite{si-gga,imma-the}}$, to be stable between the $\beta$Sn and sh phases.
However, the main features of differences in the phase transitions between the hydrostatic and uniaxial compression, which are what we would like to address in the present work, can already be demonstrated by considering the five phases included here.

\section{Possible patterns of uniaxial phase transitions}

As discussed in the paper I, the appropriate physical quantity to consider in determining the stable phases under uniaxial condition is$^{\cite{note}}$
\begin{eqnarray}
 \bar{\Delta}H & = & [(E_{2}-E_{1})+P_{av}(V_{2}-V_{1})] + (P_{z}-P_{x})[(\frac{1}{3}(V_{1}-V_{2})+\int_{1}^{2}l_{x}l_{y}dl_{z})] \\
        & \equiv & \dh + \bwbar
\end{eqnarray}
where the limits of the $l_{z}$ in the integral are the lattice constants of the crystal phases 1 and 2 under uniaxial pressures $P_{x}$ and $P_{z}$.
The internal energies (E) and volumes (V) (subscript "1" and "2" for the phases 1 and 2 respectively) are those when the phases are under the uniaxial pressures $P_{x}$ and $P_{z}$ while $P_{av}$ is the averaged pressure, i.e. $(2P_{x}+P_{z})/3$.
Phase 2 is relatively more stable than phase 1 if the quantity \bhbar is less than zero and the phase-transition pressures are those when the condition $\bhbar=0$ is met.
In addition to the \dh term, there is a \bwbar which is not present under hydrostatic condition.
The magnitude of \bwbar depends on the path of the integration and reduces to zero only under hydrostatic condition.
The physically most reasonable path for the integral is the shortest path, as discussed in paper I, and the integral can be written as
\begin{equation}
 \frac{1}{3}(a_{z,2}-a_{z,1})(A_{2} + \sqrt{A_{2}A_{1}}+A_{1})
\end{equation}
for phases considered in the present study.
By using equation (4), the quantity \bwbar can be rearranged as
\begin{equation}
 \bwbar  =   \frac{1}{3}(P_{z}-P_{x})(\sqrt{A_{2}}+\sqrt{A_{1}})(a_{z,2}\sqrt{A_{1}} - a_{z,1}\sqrt{A_{2}}). 
\end{equation}
We shall investigate the possible patterns of uniaxial phase transitions by studying the {\em ab initially} obtained phase-transition curves of $\bhbar = 0$ on the $P_{x}$-$P_{z}$ plane for any two chosen phases of the five Si structures, i.e. total ten cases will be studied.
The possible patterns of $\bhbar = 0$ will be discussed in terms of the contributions from \dh  and \bwbar respectively and analyzed using their explicit formuli (equations (2) and (5)).

All the possible patterns of the uniaxial phase-transition curves $\bhbar = 0$ are schematically plotted in Fig. 2a and 2b which correspond to the cases whose $P_{hydro}$ of the two considered phases are positive and negative respectively.
For the cases with positive $P_{hydro}$, if there exist uniaxial phase-transition pressures (both $P_{x}$ and $P_{z}$) being smaller than the $P_{hydro}$, they are said to have lower uniaxial phase-transition pressures.
These phase-transition curves have positive gradients around $P_{hydro}$ on the $P_{x}$-$P_{z}$ plane (the thick dashed and the thick dotted-dashed lines in Fig. 2a).
On the other hand, in the cases of higher uniaxial phase-transition pressures, i.e. if all the uniaxial phase-transition pressures having either $P_{x}$ or $P_{z}$ or both being larger than the $P_{hydro}$, the gradients of $\bhbar=0$ are negative around $P_{hydro}$ on the $P_{x}$-$P_{z}$ plane (the thick solid line in Fig. 2a).
For the cases with negative $P_{hydro}$, if the gradients of $\bhbar=0$ are negative around $P_{hydro}$ (the thick solid line in Fib. 2b), there is no phase transition between the two considered phases when under compression.
However, if the gradients of $\bhbar=0$ are positive around the negative $P_{hydro}$ (the thick dashed and the thick dotted-dashed lines in Fib. 2b), the phase transitions between the two considered phases occur only when under uniaxial compression, in spite of no phase transition when under hydrostatic compression.

From Eq.(2), the \dh term depends explicitly on the energy and volume differences and the averaged pressure.
If the energy and volume differences of the two considered phases depend little on the pressure anisotropy, i.e. $P_{z}-P_{x}$, the contours of constant \dh are expected to be straight lines of negative gradients on the $P_{x}$-$P_{z}$ plane.
This is exactly what was found for most of the cases we study here, except for the cases of diamond-sh, diamond-sc, $\beta$Sn-hcp and sh-sc whose magnitudes of \dh are too tiny to have  the constant-\dh contours of reasonables sizes plotted.
\textit{Under uniaxial condition, the \dh alone can not lead to lower phase-transition pressures.} 

The other contribution to \bhbar is the \bwbar of equation (5).
{\em Ab initio} calculations showed that the constant-$\bar{\Delta}W$ contours on the $P_{x}$-$P_{z}$ plane are mainly straight lines of positive gradients.
Whether the contribution of \bwbar is large enough to overbalance the negative-gradient constant-\dh contours and leads to the positive-gradient constant-\bhbar contours depends strongly on the magnitude of $(a_{z,2}\sqrt{A_{1}} - a_{z,1}\sqrt{A_{2}})$ in the \bwbar.  
We found that the zero-pressure structural parameters already give a good estimate of the importance of \bwbar.
The individual values of $a_z$ and $\sqrt{A}$ may change considerably when the phases are under pressures, however, the values of $(a_{z,2}\sqrt{A_{1}} - a_{z,1}\sqrt{A_{2}})$ remain roughly a constant.
In the cases where both considered phases are approximate isotropic structures at zero pressure, i.e. $\sqrt{A}\simeq a_{z}$ in Tab. 2, the magnitude of $(a_{z,2}\sqrt{A_{1}} - a_{z,1}\sqrt{A_{2}})$ will be tiny.
The data in Tab. 2 suggest that for the phase transitions between diamond and sh, diamond and sc, and between sh and sc, the contributions of \bwbar to the \bhbar should be small.
Our {\em ab initio} calculations indeed verify that the gradients of $\bhbar=0$ around $P_{hydro}$ for these three cases are all negative.
In Fig. 3 we present the {\em ab initially} obtained $\bhbar=0$ for the diamond-to-sh (the thick solid line) and diamond-to-sc (the thick dash line) phase transitions.
These are the two cases having higher uniaxial phase-transition pressures.
As the $P_{hydro}$ for the phase transition from sc to sh is negative (corresponding to the schematically thick solid line in Fig. 2b), the sh phase is always more stable than the sc phase under positive external pressures.
In general, the sign of $P_{hydro}$ can be decided from the zero-pressure properties of the phases as well.
For two considered phases, if the one which has larger zero-pressure volume but lower zero-pressure energy, the $P_{hydro}$ is positive.
On the other hand, if the one which has larger zero-presure volume also has larger zero-pressure energy, the $P_{hydro}$ is negative.

In the cases where, at zero pressure, the structure of one phase is very anisotropic while that of the other is approximate isotropic, the contribution of \bwbar to \bhbar will be significant, due to the component of $(a_{z,2}\sqrt{A_{1}} - a_{z,1}\sqrt{A_{2}})$ in \bwbar, and lead to positive-gradient constant-\bhbar contours.  
These include the phase transtions between diamond and $\beta$Sn, diamond and hcp, $\beta$Sn and sh, $\beta$Sn and sc, $\beta$Sn and hcp, sh and hcp, and between sc and hcp.
The sign of $P_{hydro}$ results in different patterns of $\bhbar=0$.
With important contribution from \bwbar, cases of positive $P_{hydro}$ have lower transition pressures (Fig. 2a) while for the cases of negative $P_{hydro}$ the phase transitions occur only at uniaxial condition, i.e. $P_{x}\neq P_{z}$ (Fig. 2b).
In the seven cases considered now, those with positive $P_{hydro}$ values are diamond-$\beta$Sn, diamond-hcp and $\beta$Sn-sh cases.
The {\em ab initially} obtained phase-transition curves $\bhbar=0$ for the three cases (shown in Fig. 4) confirm the conjecture that in these three cases, uniaxial compression greatly reduce the phase transition pressures.
That is, the gradients of $\bhbar=0$ for these three cases are all positive around $P_{hydro}$.
Whether the \bhbar has gradient larger or smaller than one can be decided by the relative structural parameters of the two considered phases at zero pressure.
In the cases of diamond-to-$\beta$Sn and diamond-to-hcp transitions, both the $\beta$Sn and hcp phases have larger $\sqrt{A}$ than $a_{z}$ while the diamond phase has equal magnitude of $\sqrt{A}$ and $a_{z}$ at zero pressure.
Phase transitions from diamond to either $\beta$Sn or hcp are expected to be easier if the $P_{z}$ pressure is larger than the $P_{x}$ as under this condition the dimensions of the compressed diamond phase are forced towards the anisotropic ones ($\sqrt{A}>a_{z}$) as those of $\beta$Sn or hcp.
Similarly in the case of $\beta$Sn-to-sh transition, the transition is expected to be easier if $P_{x}$ is larger than $P_{z}$ as the transition is from the anisotropic phase of $\beta$Sn ($\sqrt{A}>a_{z}$) towards the approximate isotropic sh phase.
These intuitive ideas are supported by the explicit formula of \bwbar in Eq. (5).
However, this way of thinking should not be pushed too far as the phase transition is first order and there is a discontinuity in $\sqrt{A}$, $a_{z}$, as well as the structures.  

The remaining four cases, i.e. $\beta$Sn-sc, $\beta$Sn-hcp, sh-hcp and sc-hcp which have important contribution of \bwbar and negative $P_{hydro}$, should have the $\bhbar=0$ contours look like either the thick dashed or the thick dotted-dashed line in Fig. 2b.
However, the $P_{hydro}$ for the $\beta$Sn-hcp is so large in magnitude that the \bhbar$=0$ is still out of the positive-pressure quardrant of the $P_{x}$-$P_{z}$ plane we included in our {\em ab initio} studies, i.e. in the region of $P_{x}<20GPa$ and $P_{z}<20GPa$.
For the case of $\beta$Sn and sc phases, as the transition from $\beta$Sn to sc is to transit from the phase of $\sqrt{A}>a_{z}$ to the more isotropic phase, the transition will be easier if $P_{x}>P_{z}$.
This is exactly what we obtained from {\em ab initio} calculations (the thick dotted-dashed line in Fig. 5).
For the cases of sh-to-hcp and sc-to-hcp transitions, the situations are opposite to the $\beta$Sn-to-sc transition as now the transitions are from the more isotropic phases (sh and sc) to the phase having $\sqrt{A}>a_{z}$ (hcp).
The phase transition line \bhbar$=0$ is expected to have the property of $P_{x}<P_{z}$.
In the case of sc-to-hcp, the contribution of the \dh is tiny and the \bwbar is the dominant factor, therefore the \bhbar$=0$ phase-transition line is almost parallel to the $P_{x}=P_{z}$ line.
Note that for the last three considered cases of $\beta$Sn-sc, sh-hcp and sc-hcp, although there is no phase transition at hydrostatic condition, the phases do transit when under uniaxial compression, i.e. $P_{x}\neq P_{z}$.
That is, there might be phase transitions which cannot be observed when the phases are under hydrostatic compression but can be observed when under uniaxial compression.

\section{Phase Diagram of Si}

In the previous section, we discuss the possible phase-transition patterns of Si by considering only two phases of the five strucutres at a time.
However, the phase diagram should be constructed when all the phases are taken into consideration.
With all the five structures taken into consideration, we obtain the phase diagram of Si under uniaxial compression as shown in Fig. 6.
The stable phases are diamond, $\beta$Sn and sh which are the same as those when under hydrostatic condition but stabilize at different regions of the $P_{x}$-$P_{z}$ plane.
Under hydrostatic compression, the phase transition is from diamond to $\beta$Sn on increasing pressures and to sh if the pressures are increased even further.
However, the phase diagram suggests that if the external pressures are kept under the condition of $P_{x}>P_{z}$, on increasing pressures, the phase transition should go directly from the diamond to the sh phase (e.g. the horizontal dashed arrow in Fig. 6).
Similarly, the sh-to-$\beta$Sn transition on increasing pressures, which is not possible when the phases are under hydrostatic compression, can be observed, according to the phase diagram, if the applied uniaxial pressures are varied from the condition of $P_{x}>P_{z}$ to that of $P_{x}<P_{z}$ (e.g. the vertical dashed arrow in Fig.6).

\section{Conclusion}

Based on a previously proposed thermodynamic analysis, we have studied the possible phase-transition patterns between phases by carrying out {\em ab initio} calculations for five Si phases, i.e. diamond, $\beta$Sn, sh, sc and hcp.
We have shown that it is possible to guess the phase-transition pattern of two considered phases by inspecting their structural parameters at zero pressure.
For cases with positive $P_{hydro}$, uniaxial compression can greatly reduce the phase-transition pressure if the structure of one phase is approximate isotropic while the other is very anisotropic.
For cases with negative $P_{hydro}$, it is possible to have phases, which do not undergo phase transitions when under hydrostatic compression, do undergo phase transitions when under uniaxial compression.
With all the five phases taken into consideration, the phase diagram of Si when under uniaxial compression was constructed.
It was shown that, in spite of the same stable phases were found as those under hydrostatic compression, very different phase transitions were possible depending on the conditions of the applied uniaxial pressures.

The author gratefully acknowledges W. H. Huang for invaluable discussions.
This work wass supported by the National Science Council in Taiwan.
The computer resources were partly provided by the National Center for High-Performance Computing in HsinChu, Taiwan.

\bibliographystyle{unsrt}

\begin{thebibliography}{80}
\bibitem{note}
 We have changed the definitions of \dh and \bwbar from those used in the paper I.
\end{thebibliography}

\pagebreak
\noindent {\large\bf Tables:} \\
\noindent {\large\bf Table 1.} 
The calculational details used in the present study.  See the text in section 2 for details.
\begin{center}
\begin{tabular}{cccccc} \hline\hline
                  & diamond & $\beta$-Sn  & sh  & sc & hcp  \\  \hline
8-Si unit cell & tetragonal & tetragonal & $2\times2\times2$ & $2\times2\times2$ & $2\times2\times1$ \\
k-point sets     & (4 4 4) & (8 8 20) & (9 9 9) & (10 10 10)  & (7 7 17) \\
$a_{lateral}$ ($\AA$) & 4.60$\sim$5.80 & 6.30$\sim$6.95 & 4.65$\sim$5.40 & 4.50$\sim$5.20  & 5.70$\sim$7.70 \\
$a_{z}$ ($\AA$)       & 4.60$\sim$6.20 & 2.40$\sim$2.90 & 4.30$\sim$5.35 & 4.50$\sim$5.20  & 2.10$\sim$3.70  \\  \hline\hline
\end{tabular}
\end{center}

\noindent {\large\bf Table 2.} 
The calculated structural parameters and total energies of the five phases at zero pressure.
\begin{center}
\begin{tabular}{cccccc} \hline\hline
                  & diamond & $\beta$-Sn  & sh  & sc & hcp  \\  \hline
$a_{lateral}$ ($\AA$)  & 5.459 & 6.815 & 5.295 & 5.067  & 7.096 \\
$\sqrt{A}$ ($\AA$)     & 5.459 & 6.815 & 4.927 & 5.067  & 6.603 \\
$a_{z}$ ($\AA$)   & 5.459 & 2.650 & 4.991 & 5.067  & 3.002 \\
V ($\AA^{3}/$Si)  & 20.334 & 15.385 & 15.147 & 16.257  & 16.363  \\
E (eV/8Si)        & -43.546 & -40.934 & -40.777 & -40.562  & -40.021 \\  \hline\hline
\end{tabular}
\end{center}

\pagebreak
{\large \bf Figure Captions:}\\
{\bf Figure 1.} \\
Energy of the five Si phases, i.e. diamond, $\beta$Sn, sh, sc and hcp, plotted versus volume.  
\\ 
{\bf Figure 2.}\\
Schematic plots showing possible patterns of phase-transition curves (the thick lines) : (a) cases with $P_{hydro}>0$ and (b) cases with $P_{hydro}<0$.
The phase-transition curves correspond to the $\bhbar(P_{x},P_{z})=0$ contours.
The thin solid lines are the line of $P_{x}=P_{z}$ and the thin dashed lines are to guide the eyes to show the hydrostatic phase-transition pressures $P_{hydro}$.
\\ 
{\bf Figure 3.} \\
The {\em ab initially} obtained phase-transition curves, i.e. the $\bhbar(P_{x},P_{z})=0$ contour, for the diamond-to-sh (the thick solid line) and diamond-to-sc (the thick dashed line) phase transitions.
In both cases, the diamond structure stabilizes at the lower and left-handed regions of the phase-transition lines. 
The thin solid line, i.e. the line of $P_{x}=P_{z}$, corresponds to the hydrostatic condition.
\\
{\bf Figure 4.} \\
The {\em ab initially} obtained phase-transition curves for the diamond-to-$\beta$Sn (the thick dashed line), diamond-to-hcp (the thick dotted-dashed line), and the $\beta$Sn-to-sh (the thick solid line) transitions.
In both the diamond-$\beta$Sn and diamond-hcp cases, the stable phases of the lower and right-handed regions of the transition lines are the diamond structure. 
In the $\beta$Sn-sh case, the stable phase of the lower and right-handed region of the transition line is the sh structure.
The thin solid line corresponds to the hydrostatic condition, i.e. $P_{x}=P_{z}$.
\\
{\bf Figure 5.} \\
The {\em ab initially} obtained phase-transition curves for the $\beta$Sn-to-sc (the thick dotted-dashed line), sh-to-hcp (the thick dashed line), and the sc-to-hcp (the thick solid line) transitions.
The stable phases in the lower and right-handed regions of the transition lines are the sc, sh and the sc structure for the $\beta$Sn-sc, sh-hcp, and sc-hcp cases respectively. 
The thin solid line corresponds to the hydrostatic condition, i.e. $P_{x}=P_{z}$.
\\
{\bf Figure 6.} \\
The phase diagram of Si under uniaxial compression of $P_{x}<20GPa$ and $P_{z}<20GPa$ when the five phases, i.e. diamond, $\beta$Sn, sh, sc, and hcp, were taken into consideration.
The phase boundary between the $\beta$Sn and sh phases is denoted by the dashed line in order to show its difference from the hydrostatic line of $P_{x}=P_{z}$ (the thin solid line).
The horizontal and vertical arrows show the possible phase transitions which are not possible when under hydrostatic compression, i.e. along the thin solid line of $P_{x}=P_{z}$.
\\

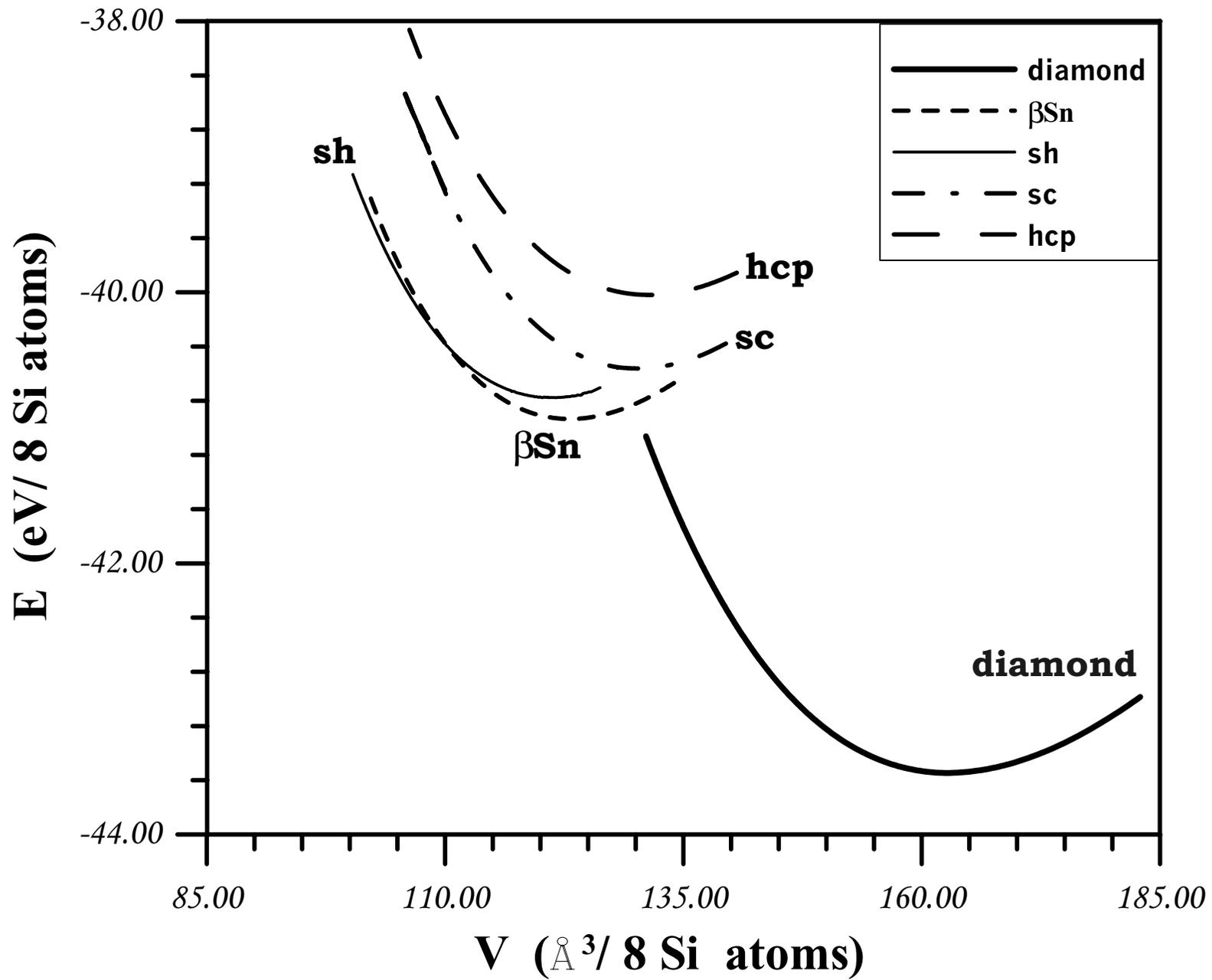
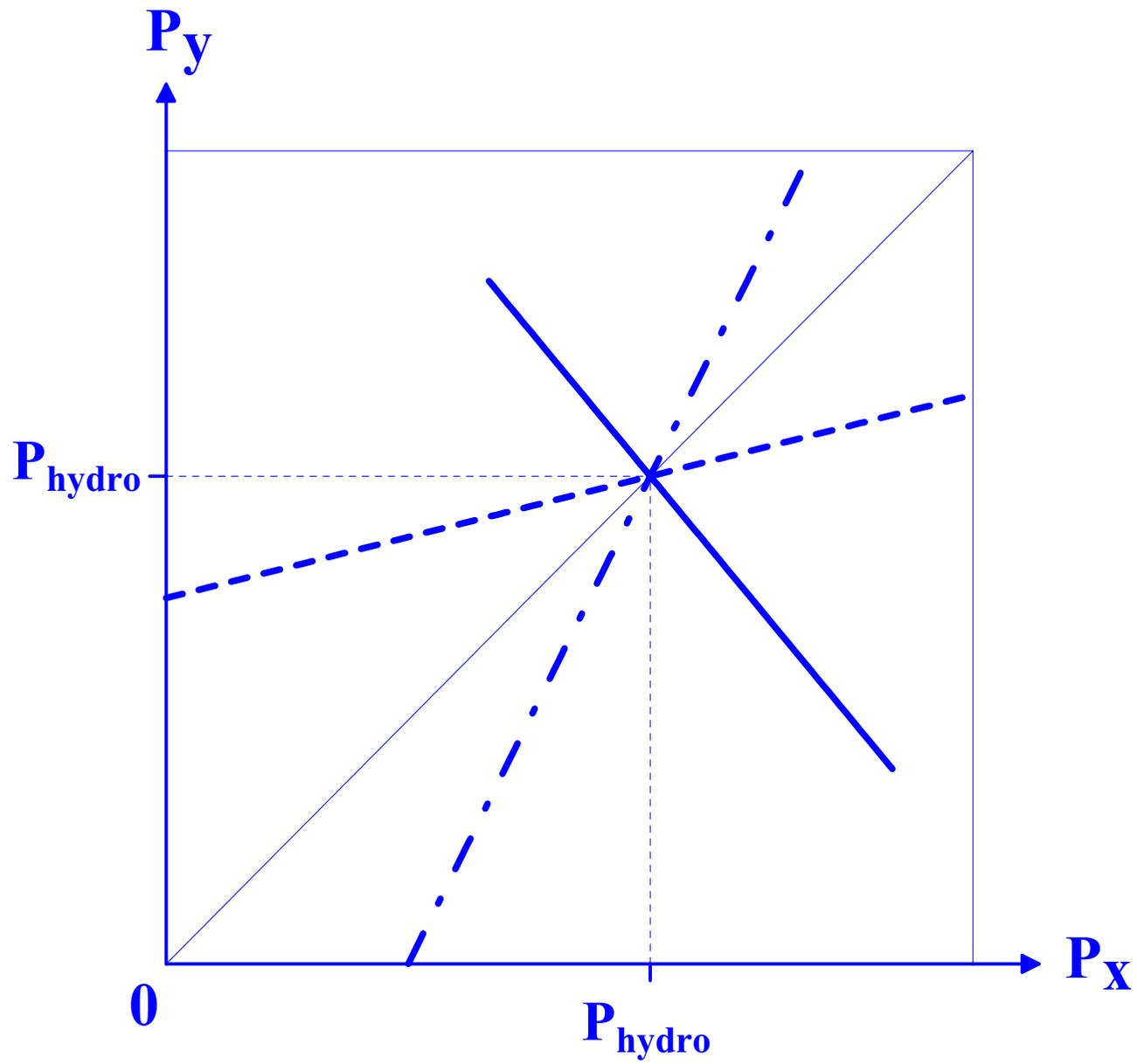
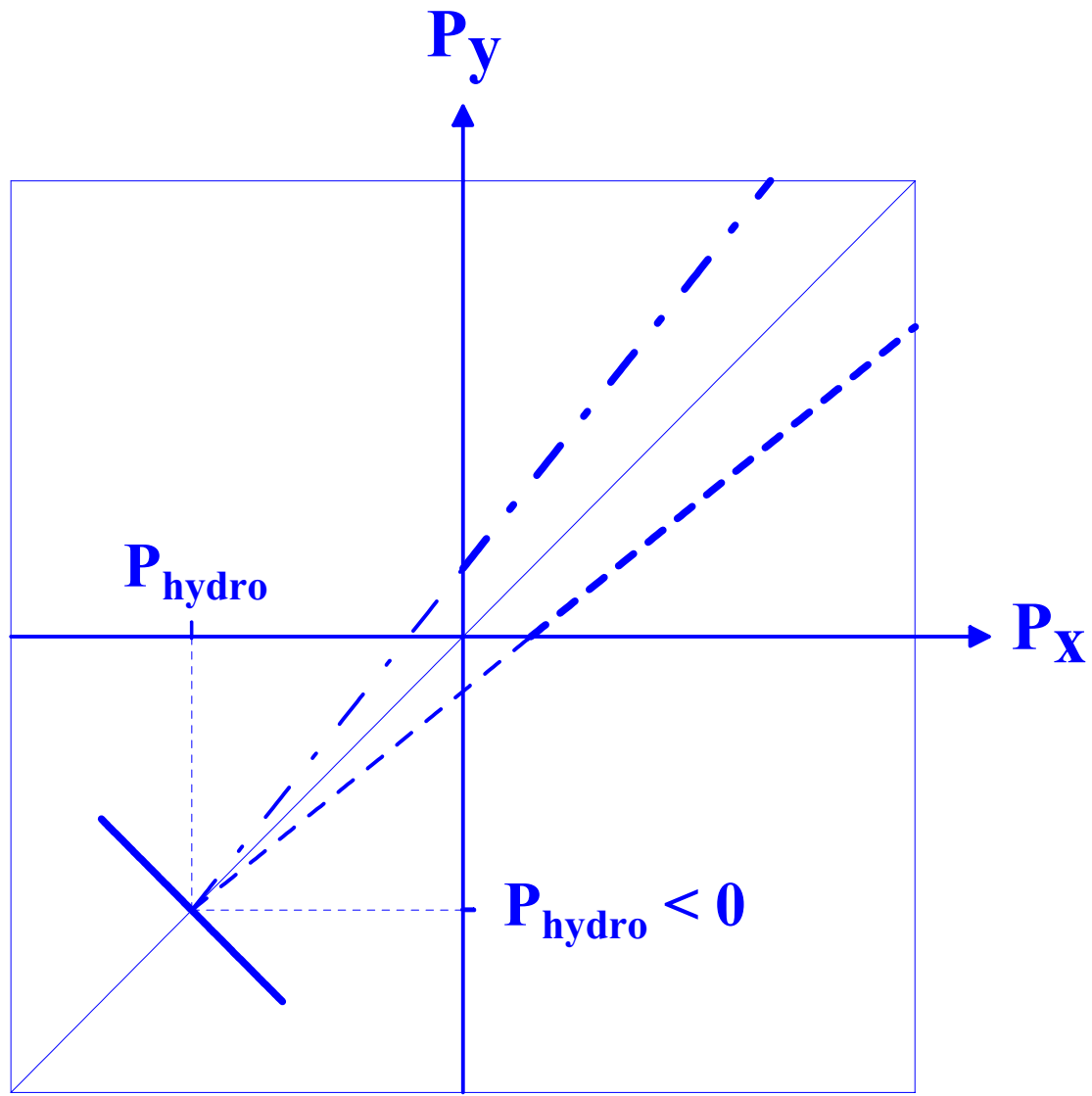
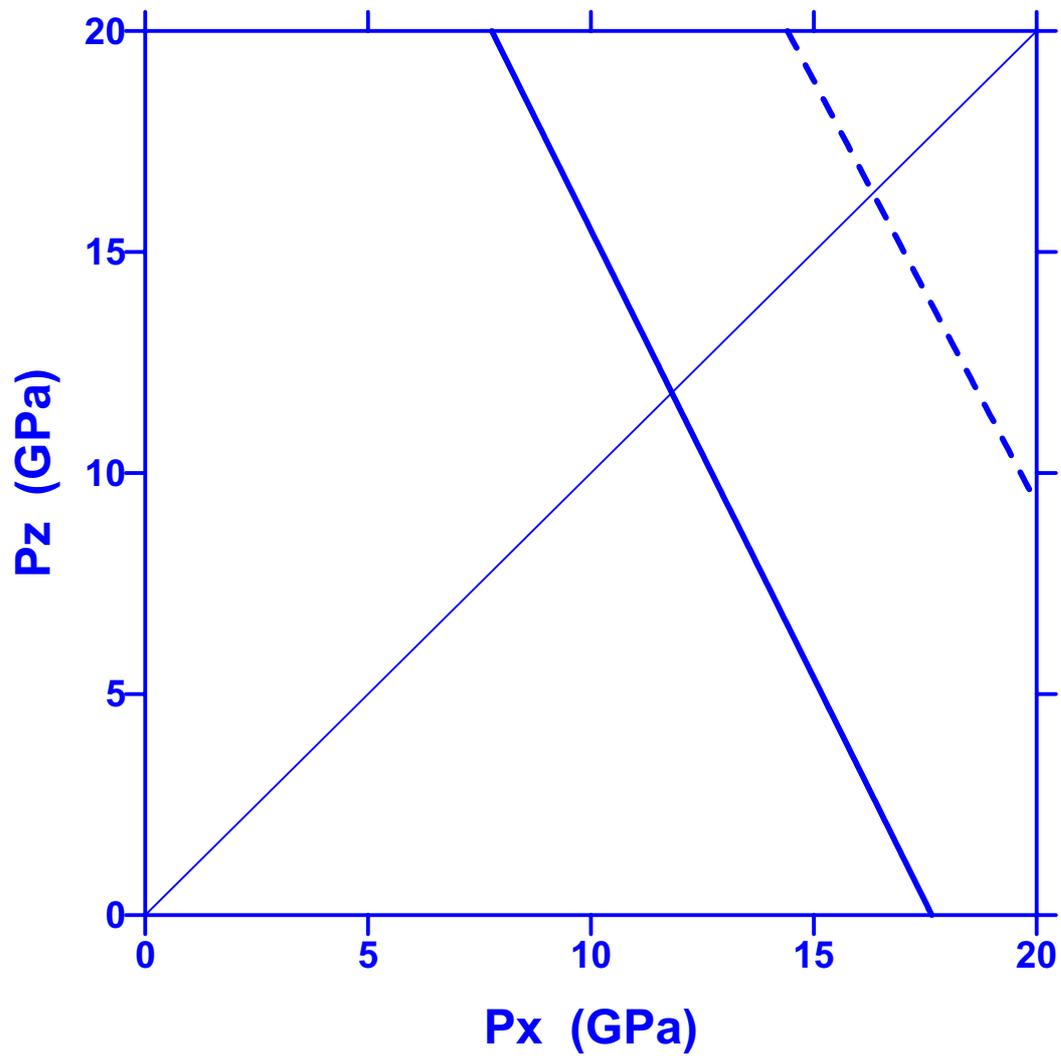
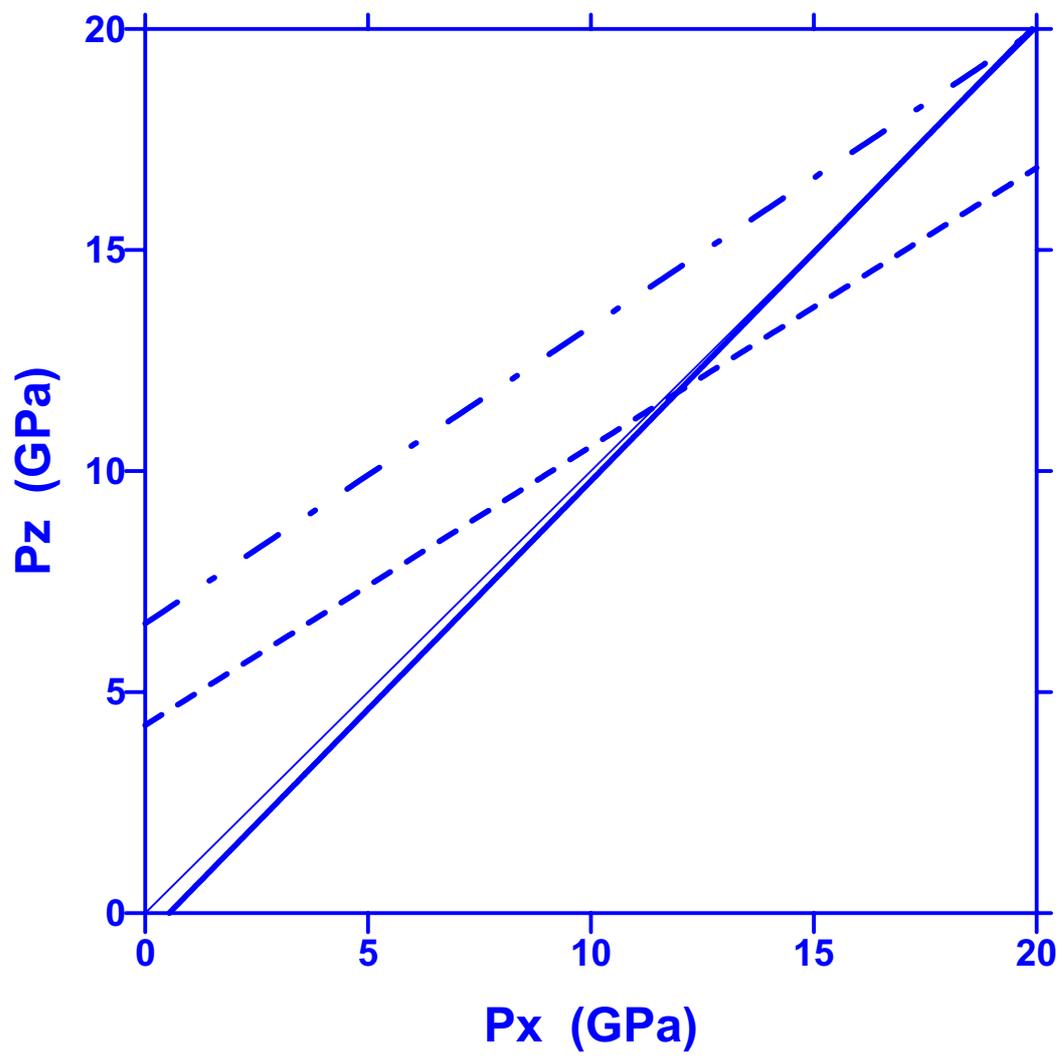
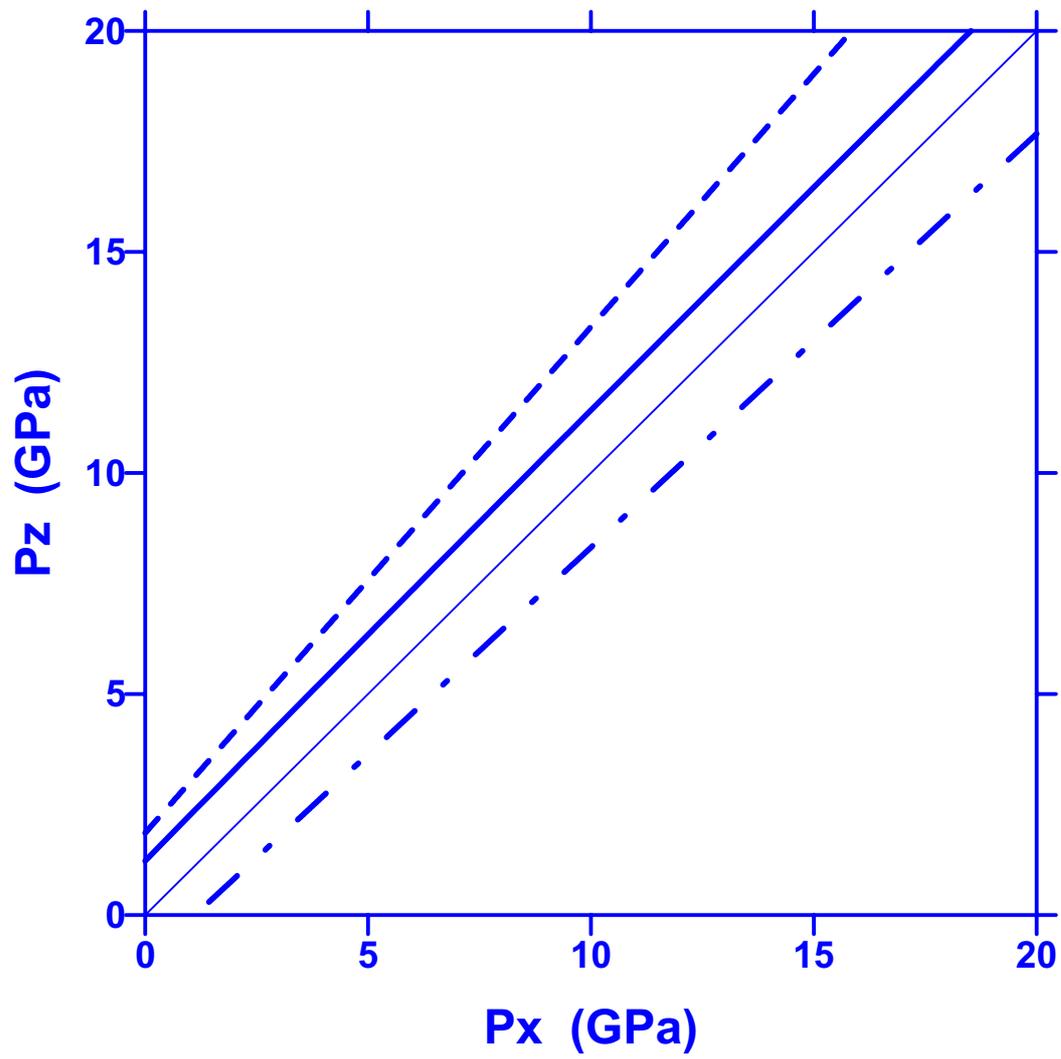
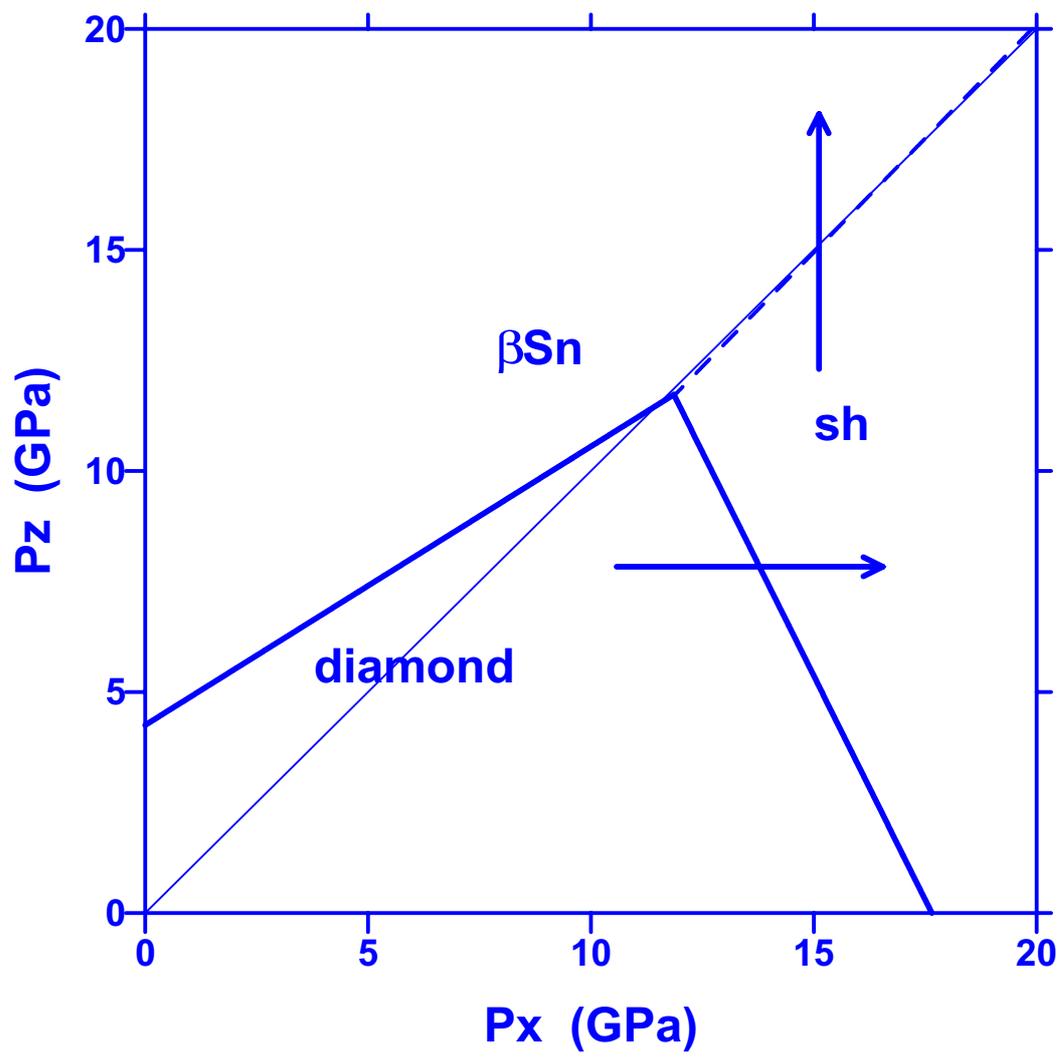
\end{document}